\documentclass[a4paper]{jpconf}
\usepackage{wrapfig}
\usepackage[colorlinks,citecolor=blue,linkcolor=red]{hyperref}
\usepackage{amsfonts,amssymb,amscd,amsmath}
\usepackage{amssymb}
\usepackage{amsmath}
\usepackage{graphicx}
\usepackage{graphics}
\usepackage{pifont}
\usepackage{hyperref}

\def\dis{\displaystyle}
\def\le{\left(}
\def\ri{\right)}
\def\no{\nonumber}

\usepackage{graphicx}
\begin{document}
\title{Algorithm to find an all-order in the running coupling solution to an equation of the DGLAP type}

\author{Igor Kondrashuk}

\address{Grupo de Matem\'atica Aplicada, Departamento de Ciencias B\'asicas, Univerdidad del B\'io-B\'io, Campus Fernando May, Casilla 447, Chill\'an, Chile}

\ead{igor.kondrashuk@gmail.com}

\begin{abstract}
We propose an algorithm to find a solution to an integro-differential equation of the DGLAP type for all the
orders in the running coupling $\alpha$ with splitting functions given at a fixed order in $\alpha.$ Complex analysis is
significantly used in the construction of the algorithm, we found a simpler way to calculate the involved integrals
over contours in the complex planes than by any of the methods known at present.
\end{abstract}

\section{Introduction}

The structure functions of proton may be measured experimentally in deep inelastic scattering processes \cite{Bjorken:1968dy}. Gribov and Lipatov studied these processes in QED and 
found that these structure functions satisfy certain integro-differential equations \cite{Gribov:1972ri,Gribov:1972rt,Lipatov:1974qm}. The renormalization group equations (RGEs) for the coefficient functions 
of operator product expansion have been obtained in the Nobel prize paper \cite{Gross:1974cs} which marked the discovery of QCD.  Then, Altarelli and Parisi wrote these
RGEs for the coefficient functions 
of Wilson operator product expansion in an integro-differential form and interpreted them as integro-differential equations (IDEs) for the parton distribution functions (PDFs) \cite{Altarelli:1977zs}.
Similar IDEs were written at the same time by Dokshitzer \cite{Dokshitzer:1977sg} who developed to the QCD case the Gribov and Lipatov approach used in QED.
These IDEs became known as Dokshitzer-Gribov-Lipatov-Altarelli-Parisi equation (also known as DGLAP equation).

It is known that two steps are necessary in order to solve an IDE  of the DGLAP type. The first step is to take the Mellin moment of both the sides of the DGLAP IDE with respect to Bjorken $x.$
This transforms the IDE of the DGLAP-type  to a first-order differential equation with respect to the scale of momentum transfer and to solve this 
differential equation for the Mellin moment.  The fully analytic solutions of the non-singlet and singlet evolution equations are known since the late
nineties at the next-to-next to leading order with small $x$ resummations included. In \cite{Blumlein:1997em,Diemoz:1987xu} the   
evolution operator in the case of the running coupling has been constructed. These equations may be solved, the recent developments may be found in  \cite{Ablinger:2018zwz}.  
On the other hand, Mellin space solutions even to NNLO with the running coupling exist, and have been worked out in several numerical codes, for example \cite{Vogt:2004ns}.  There are various later numerical software 
packages too [see e.g. the citations of \cite{Vogt:2004ns}].

Suppose we know the solution to the corresponding differential equation. 
This talk is devoted to the second step which consists in the inverse Mellin transformation to the Bjorken $x$-space. The traditional way is to decompose the exponential function  in terms of the coupling 
(fixed or running) which we evaluate then at the residues on the complex plane of the  Mellin moment, see for example \cite{Kosower:1997vj,Altarelli:2000dw,Ball:2005mj}. 
In such a way we calculate an infinite series 
of the inverse transformations, each one is for every order of the coupling. At the lowest order in the running coupling the calculation may  be done analytically, however even at this level  a lot of work is required
when the real QCD 
case is considered instead of toy models.  
There are different software packages available to do all these steps analytically, at least at the leading order. 
A numerical software package ``PartonEvolution'' has been developed in \cite{Weinzierl:2002mv}.  Another numerical package QCDNUM has been created later \cite{Botje:2010ay,Botje:2016wbq}.

At higher orders new advanced analytical software tools exist. They are based on using  concepts from algebraic geometry like a shuffle product \cite{Ablinger:2010kw,Ablinger:2013hcp}. 
Harmonic polylogarithms are described in  \cite{Remiddi:1999ew}. 
Here we propose an alternative way which reduces  evaluation  of the inverse Mellin transformation to calculation of the inverse Laplace transformations of Jacobians corresponding  to certain complex diffeomorphisms. 
These diffeomorphisms may be performed  in the complex plane of the Mellin moment.  These complex mappings make the structure of the integrands uniform reducing it in many  of the cases to the standard tables like 
\cite{gradshteyn-2015a}.

We have shown in Ref.\cite{Alvarez:2019eaa} that the table integrals which appear in the result of this complex mapping may  further be transformed to Barnes integrals. The latters are the
standard representations for generalized hypergeometric functions. In order to transform the original contour integral to the Barnes integrals 
we curve the vertical line of the contour integral in the inverse Laplace transformation to the shape of the Hankel contour and then calculate the corresponding Hankel contour integral  of the product of the 
exponential function, typical for the inverse Laplace transformation,  with a power of the variable from the Jacobian of this complex map and get in such a way the gamma function in the denominator of the Barnes integral. 
This is the reason why we prefer to have the inverse Laplace transformation at the intermediate step instead the inverse transformation of the Mellin moments.

The structure in the Mellin space in Ref. \cite{Blumlein:1997em} suggests that analytic inversions of some evolution kernels to $x$-space usually will need the introduction of new special functions 
on just the perturbative side both in the cases of the fixed and of the running coupling.  In this talk we suggest that the origin of these new functions may be classified by  certain diffeomorphisms in the complex plane 
of the Mellin moment.

There is another important point. Physically, to introduce these functions is, however, not very useful either, since
the parton densities have still to be Mellin-folded with the shape of PDFs  at the starting scale $Q_0^2$  in the $x$-space,  and one has no a priori shape parameterization for those.
In general, we have to parametrize a distribution at some scale and evolve this distribution according to the DGLAP IDE.  
One of the parameterizations is a Bernoulli parametrization \cite{Ball:2016spl}  at the scale $Q_0^2,$  
see for example \cite{Alekhin:2003ev,Blumlein:2006ws,Alekhin:2012ig}.  Instead of one Beta function a combination of them may be taken \cite{Alekhin:2003ev}.

The convolution with Bernoulli distribution usually may be done numerically. This is the common philosophy that it is better 
to parameterize the PDFs in the Mellin space and to perform the one needed final numerical integral as the inversion integral itself.
This is widely agreed in the community that one integral in all cases has to be numerical to allow for flexible enough inputs at
the starting scale.  As we have explained above we do not need this last integral to do numerically.

The complete form of the DGLAP equations for PDFs in QCD may be found in the original papers \cite{Altarelli:1977zs,Dokshitzer:1977sg}.
\begin{eqnarray}
\label{differences} u\frac{d}{du}\Delta_{ij}(x,u) &=& \frac{\alpha(u)}{2\pi}\int_x^1\frac{d y}{y}\Delta_{ij}(y,u)P_{qq}\left(\frac{x}{y}\right) \\
\label{SD} u\frac{d}{du} \Sigma (x,u) &=& \frac{\alpha(u)}{2\pi}\int_x^1\frac{dy}{y}\left[\Sigma(y,u)P_{qq}\left(\frac{x}{y}\right) + (2f)G(y,u)P_{qG}\left(\frac{x}{y}\right)\right]
\end{eqnarray}
\begin{eqnarray}
\label{GD} u\frac{d}{du} G(x,u)  &=& \frac{\alpha(u)}{2\pi}\int_x^1\frac{d y}{y}\left[\Sigma(y,u)P_{Gq}\left(\frac{x}{y}\right) +  G(y,u)P_{GG}\left(\frac{x}{y}\right)\right] 
\end{eqnarray}
with $\Delta_{ij}(x,u) = q_i(x,u) - q_j(x,u)$ and $\Sigma(x,u)= \sum_i \left[q_i(x,u) + \overline{q}_i(x,u)\right],$ where  $\Delta_{ij}(x,u)$ is non-singlet distribution function, $\Sigma(x,u)$ is a singlet 
distribution function, $G(x,u)$ is a gluon distribution function, $P_{qq},P_{qG},P_{GG}$ are the splitting functions, which may be found from anomalous dimensions of operators in QCD \cite{Vogt:2004mw,Moch:2004pa}, 
$u \equiv Q^2/\mu^2,$ $Q^2$ is the momentum transfer,  $\mu^2$ is a scale, $\alpha(u)$ is the running coupling.

\section{Analytical solution to DGLAP equation}

In this section we consider an analytical solution to the evolution equation, using a very simple toy model. To begin, an 
IDE of DGLAP-type in which the splitting function $P(x)$  participates may be written as  
\begin{eqnarray} \label{DGLAP-1} 
u\frac{d}{du} G(x,u) =  \frac{\alpha(u)}{2\pi}\int_x^1\frac{dy}{y}~G(y,u)P\left(\frac{x}{y}\right).
\end{eqnarray}
The standard way is to take Mellin $N$-moment of both the parts of this equation and to obtain the relation 
\begin{eqnarray} \label{DGLAP-2}
u\frac{d}{du} \int_0^{1}dx~x^{N-1}G(x,u)  =  \frac{\alpha(u)}{2\pi}\int_0^{1}dx~ x^{N-1}\int_x^1\frac{dy}{y}~G(y,u)P\left(\frac{x}{y}\right)  \no \\
=  \frac{\alpha(u)}{2\pi}\int_0^{1}dy~y^{N-1}G(y,u) \int_0^1 dx ~x^{N-1} P\left(x\right) = \frac{\alpha(u)}{2\pi} ~\gamma(N,\alpha(u)) ~M[G(y,u),y](N),
\end{eqnarray}
in which the anomalous dimension $\gamma(N,\alpha(u))$ is defined as $\int_0^1 dx ~x^{N-1} P\left(x\right) = \gamma(N,\alpha(u)),$ and 
$M[G(x,u),x](N)$ is the Mellin moment of $G(x,u)$ with respect to the variable $x.$ In the limit of small $x$ Eq.(\ref{DGLAP-2}) is an approximation to the system of Eqs.(\ref{differences}-\ref{GD})
because the gluon distribution becomes dominant in this limit.  

The gluon distribution $G(x,u)$ is called integrated gluon distribution. It is related to unintegrated gluon distribution $\varphi(x,k_\perp^2)$ by the relation 
\begin{eqnarray*}
G\le x,\frac{Q^2}{\mu^2}\ri = \int_0^{Q^2}dk_\perp^2 \varphi\le x,\frac{k_\perp^2}{\mu^2}\ri. 
\end{eqnarray*}
It appears, it is always possible to construct from $\varphi(x,k_\perp^2)$ a function which satisfies the same DGLAP equations as well as the integrated gluon distributions do. 
For example, in case when the coupling constant does not run, we have the same DGLAP equation for  $ \phi\le N,\frac{Q^2}{\mu^2}\ri \equiv Q^2\varphi\le N,\frac{Q^2}{\mu^2}\ri,$
that is \cite{Alvarez:2016juq}
\begin{eqnarray*}
Q^2\frac{d}{dQ^2} Q^2\varphi\le N,\frac{Q^2}{\mu^2}\ri = \frac{\alpha}{2\pi}\gamma(N,\alpha)   Q^2  \varphi\le N,\frac{Q^2}{\mu^2}\ri.
\end{eqnarray*}
We may substitute the inverse Mellin transform  $\int_{a-i\infty}^{a+i\infty}~dN x^{-N} \phi(N,u)$   into this DGLAP equation and  obtain the equation  
\begin{eqnarray*}
\int_{a-i\infty}^{a+i\infty}~dN x^{-N} \phi_1(N)u^{{\frac{\alpha}{2\pi}\gamma(N,\alpha)}}\left[\gamma(N,\alpha) - 
x^N\int_x^1\frac{d y}{y} y^{-N}  P_{GG}\le \frac{x}{y},\alpha\ri \right] = 0,
\end{eqnarray*}
which may be treated as a self-consistency condition for the shape $\phi_1(N)$ \cite{Alvarez:2016juq}.

If we leave only one term in the splitting function, the solution for unintegrated gluon distribution function in the small $x$ limit is simple \cite{Blumlein:1995eu}.  For example, 
\begin{eqnarray} \label{blum}  
P_{GG}(z,\alpha) =   2z.
\end{eqnarray}
In such a case, the self-consistency condition above is  $\dis{\int_{a-i\infty}^{a+i\infty} ~dN \frac{\phi_1(N)u^{{\frac{\alpha}{2\pi}\gamma(N,\alpha)}}}{N+1} = 0},$
the anomalous dimension is  $\dis{\gamma(N,\alpha)  = \frac{2}{N+1}},$ and if we choose  $\phi_1(N)  =  1/(N+1)$ to fulfill the self-consistency condition, we obtain the
Bessel function in the result of the inverse Mellin transformation 
\begin{eqnarray*} 
\phi(x,u) = \int_{-1+\delta-i\infty}^{-1+\delta+i\infty}~dN \frac{x^{-N}}{N+1} u^{ { \frac{\alpha}{\pi} \frac{1}{N+1} }} = 
x\sum_{k=0}^{\infty}\frac{1}{k!}\le\frac{\alpha}{\pi}\ln{u}\ri^k \int_{-1+\delta-i\infty}^{-1+\delta+i\infty}~dN \frac{x^{-N-1}}{(N+1)^{k+1}} \no \\
= x\sum_{k=0}^{\infty}\frac{1}{k!}\le\frac{\alpha}{\pi}\ln{u}\ri^k \frac{(-\ln{x})^{k}}{k!} 
= xI_0\le 2\sqrt{\frac{\alpha}{\pi}\ln{u}\ln{\frac{1}{x}}} \ri,
\end{eqnarray*}

\section{Duality between DGLAP and BFKL via a complex diffeomorphism}

Originally, our idea to use complex diffeomorphisms in order to make structures of the integrands uniform has appeared when we tried to find  a new approach to a duality of two IDEs,
the DGLAP and the BFKL \cite{Lipatov:1976zz,Fadin:1975cb,Kuraev:1976ge,Kuraev:1977fs,Balitsky:1978ic}, for unintegrated PDFs. The issue of this duality  was developed in Refs. 
\cite{Altarelli:2000dw,Ball:2005mj} in the late nineties and at the beginning of the years two thousand 
and served to evaluate  
enhancement of small $x$ contributions to PDFs.
The small $x$ contributions could be resummed via BFKL equation \cite{Catani:1994sq}. At the same time, ways to solve the BFKL IDE with the running coupling 
may further be searched for. Complex diffeomorphisms in the plane of the Mellin moment may appear to be a powerful tool  here. In this section we analyze a duality relation via complex mappings 
and in the next section we apply the same trick to reduce calculation of PDFs to the inverse Laplace transformation of the Jacobians for diffeomorphisms.  
The main trick is that we obtain the BFKL equation from the DGLAP equation by complex mappings.

We take a result from the previous section,  $\dis{\phi(x,u) =  \int_{-1+\delta-i\infty}^{-1+\delta+i\infty}~dN \frac{x^{-N}}{N+1} u^{ \dis{ \frac{1}{N+1}}}},$  and make a complex diffeomorphism  
in the complex plane of the variable $N$ and go to another variable $M$ which is related to the original variable $N$ by the duality transformation $\gamma(N) = M$  in the integral.  
If we define the inverse function $\chi(M) \equiv N,$ then we model the duality condition  $\chi(\gamma(N)) =  N$ and $\gamma(\chi(M)) =  M.$  To our knowledge, this duality condition for the first time 
appeared in the paper of Catani and Hautman in 1994, Ref.\cite{Catani:1994sq}. In this example we may write explicitly,   
\begin{eqnarray} 
\label{complex diffeo} \gamma(N) =  \frac{1}{N+1} = M \Rightarrow  N = \frac{1}{M} -1 \equiv \chi(M), \\  
\label{inverse} \chi(\gamma(N)) = \frac{1}{\gamma(N)} -1 = (N+1) -1 = N, ~~~ \gamma(\chi(M)) = \frac{1}{\chi(M) +1} =  \frac{1}{\left(\displaystyle{\frac{1}{M}-1}\right)  +1} = M, 
\end{eqnarray}
and make the corresponding diffeomorphism in the complex plane of the variable $N$
\begin{eqnarray*} 
\phi(x,u) =  \int_{-1+\delta-i\infty}^{-1+\delta+i\infty}~dN \frac{x^{-N}}{N+1} u^{ \displaystyle{ \frac{1}{N+1} }} =  \oint_{C}~dM N'(M) \frac{x^{-\chi(M)}} {N+1} u^M = - \oint_{C}~dM \frac{x^{-\chi(M)} }{M} u^M 
\end{eqnarray*}
\begin{eqnarray*} 
= -\oint_{C}~dM \frac{x^{{1-1/M}}}{M} u^M =  - x\oint_{C}~dM \frac{x^{-1/M}}{M} u^M = - \sum_{k=0}^{\infty}\frac{1}{k!}\le -\ln{x}\ri^k \oint_C~dM \frac{u^M}{M^{k+1}}  \\
= x\sum_{k=0}^{\infty}\frac{1}{k!}\le-\ln{x}\ri^k \frac{(\ln{u})^{k}}{k!} = x\sum_{k=0}^{\infty}\frac{(-1)^{k}}{(k!)^2} \le\ln{u}\ln{x}\ri^k = xI_0\le 2\sqrt{\ln{u}\ln{\frac{1}{x}}} \ri.
\end{eqnarray*}
We have reproduced the same Bessel function, as it should be because any integral does not depend on any change of the variable. We may write the gluon unintegrated distribution function in a dual form 
\begin{eqnarray*} 
\phi(x,u) =  \oint_{C_1}~dN \phi(N,u) x^{-N} =    \oint_{C_2}~dM \phi(x,M) u^M = xI_0\le 2\sqrt{\ln{u}\ln{\frac{1}{x}}} \ri.
\end{eqnarray*}
The contour $C_2$ is an image of the contour $C_1$ with respect to the mapping in the complex plane of the Mellin moment.  The functions $\phi(N,u)$ and $\phi(x,M)$ satisfy to the different differential equations, 
\begin{eqnarray*} 
u \frac{d}{d u} \phi \le N, u\ri   = \gamma(N,\alpha) \phi\le N, u\ri, ~~~ x \frac{d}{d x} \phi \le x, M\ri   =  -\chi(M,\alpha) \phi\le x, M\ri. 
\end{eqnarray*}
The solutions to these equations are power-like functions, $\phi \le N, u \ri = \phi_1(N) u^{\gamma(N)}, ~~ \phi \le x, M \ri = \phi_2(M) x^{-\chi(M)},$ in which the functions  $\phi_1(N)$ and   $\phi_2(M)$ 
are related by the complex diffeomorphism (\ref{complex diffeo}) which is governed by the duality condition $\chi(\gamma(N)) =  N.$ The first equation is produced by the DGLAP IDE after taking the Mellin moment 
with respect to the variable $x.$ The second equation is produced by the BFKL IDE. After making the inverse Mellin transformation in the $N$ and $M$ planes, 
this couple of differential equations becomes a couple of dual IDEs, which are the DGLAP and the BFKL equations. This couple of IDEs is called dual because they contain the 
same information about the unintegrated gluon distribution function.

In this simple model (\ref{blum}) the anomalous dimension  $\gamma(N)$ in Eq. (\ref{complex diffeo}) is a ratio of linear functions of the Mellin moment variable $N.$ As the result, the function $\chi(M)$ 
inverse to the function $\gamma(N)$ may be found by the  linear fractional transformation  (\ref{complex diffeo}). However, in the real world models like QCD to find the inverse  $\chi(M)$ function is not a simple task 
because the Euler $\psi$ function is involved already at the leading logarithm approximation (LLA)  \cite{Altarelli:2000dw,Ioffe:2010zz}. The function inverse to the  Euler $\psi$ function does not have explicit mathematical representation
however, there are several computer algorithms to evaluate it.  In the LLA 
the function $\chi(M)$ contains the combination $\psi(M) + \psi(1-M)$ that may be decomposed in terms of powers of $M$ with coefficients given by the values of Riemann zeta function
at the integer positive numbers \cite{Altarelli:2000dw,Ioffe:2010zz}. In Eq. (9.168) of the textbook \cite{Ioffe:2010zz} the solution to the BFKL equation for the Mellin moment (with respect to the Bjorken variable $x$) 
of the Green function of two interacting  reggeized gluons 
has been written in the form of a contour integral
corresponding to the inverse Mellin transformation with respect to a ratio of two squared momenta.  
The combination $N - \chi(M)$ (up to coefficients and with other letters used to denote the variables) appears in the denominator in this Eq. (9.168) 
of the contour integral of this inverse Mellin transformation. In order to take this contour integral via the residue calculus by Cauchy integral formula 
we need to find the poles in the complex plane of the variable $M$ produced by this expression and this results in  
an infinite set of terms because the structure of the BFKL eigenvalue $\chi(M)$ is not so simple like it is in the model (\ref{blum}). This task to make the residue calculus in this contour integral  
is equivalent to the search for the inverse  $\chi(M)$ function.
The main contribution to the Green function of two interacting reggeized gluons comes from the rightmost residue of the denominator $N - \chi(M).$   This 
contribution from pomeron is represented by a power-like function (9.169) from Ref.\cite{Ioffe:2010zz}. The power of this function is obtained from the rightmost residue and is given by Eq. (9.171) of \cite{Ioffe:2010zz}. 
The calculation of the next-to-leading contributions to the Green function of two interacting reggeized gluons from the BFKL pomeron due to the residue calculus in Eq. (9.168) of the textbook \cite{Ioffe:2010zz} 
may appear to be more advanced in ${\cal N} = 4$  SYM in comparison with QCD because in this maximally supersymmetric Yang-Mills theory  the corrections to the BFKL eigenvalue posses  
the property of maximal transcendentality  \cite{Kotikov:2000pm} which simplifies their structure.  This NLLA to the BFKL eigenvalue has been found by Kotikov and Lipatov 
by direct calculation in Ref. \cite{Kotikov:2000pm} where they generalized to the ${\cal N} = 4$  SYM the result of Ref. \cite{Fadin:1998py} in which the 
NLLA to the BFKL kernel was found for the QCD case. The results of Ref. \cite{Kotikov:2000pm} were reproduced 
by Dokshitzer and Marchesini with help of the reciprocity relation \cite{Dokshitzer:2006nm}. The NNLLA to  the BFKL eigenvalue has been found 
in Ref. \cite{Velizhanin:2015xsa} in which an algorithm based on a basis constructed from harmonic sums and their analytic continuations has been proposed.  
This algorithm of Ref. \cite{Velizhanin:2015xsa} treats the task to find the function inverse to the function $\chi(M).$

\section{Inverse Laplace transformation}

By using the same approach based on a change of variables in the complex plane we did in  the previous section in order to relate the DGLAP and the BFKL IDEs in case of the fixed coupling,  
we represent the inverse Mellin transformation in a form in which we may identify the inverse Laplace transformation. A part of these integrals may be found in Ref. \cite{gradshteyn-2015a}, for example.

Now we use the experience obtained in the previous section and get the same gluon unintegrated distribution function   $xI_0\le 2\sqrt{\ln{u}\ln{\frac{1}{x}}}\ri$ from another complex mapping 
which has nothing to do with the duality between the DGLAP and the BFKL IDEs, 
\begin{eqnarray*} 
\phi(x,u) =  \int_{-1+\delta-i\infty}^{-1+\delta+i\infty}~dN \frac{x^{-N}}{N+1} u^{ \displaystyle{ \frac{1}{N+1} }} = 
\int_{-1+\delta-i\infty}^{-1+\delta+i\infty}~dN \frac{e^{-N\ln{x} + \displaystyle{ \frac{\ln{u}}{N+1}}  }}{N+1} = \\
\oint_C~dM N'(M)\frac{e^{M \sqrt{  \ln{\frac{1}{x}}\ln{u}}} } {N(M)+1} =  xI_0\le 2\sqrt{\ln{u}\ln{\frac{1}{x}}} \ri, 
\end{eqnarray*}
where the complex variable $M$ is defined  by mapping  $M(N)\sqrt{  \ln{\frac{1}{x}}\ln{u}}   = -N\ln{x} + \displaystyle{ \frac{\ln{u}}{N+1}}.$
As we may see, the result obtained for the PDF appeared to be the inverse Laplace transformation of the Jacobian of the diffeomorphic mapping in the complex plane. 
This calculation has been written in Ref. \cite{Alvarez:2019eaa} in detail.

We have mentioned in the Introduction that the transformation of the original contour integral to the Barnes integrals requires 
the representation in the form of the Laplace inverse transformation at the intermediate step because we need to deform the contour of this inverse transformation to the shape of the Hankel contour in order to take one 
of the contour integrals off and obtain the Euler gamma function in the denominator.  Indeed, in Ref. \cite{Alvarez:2019eaa} we started with the contour integral which solves the DGLAP equation. Basically, this contour is a vertical line which separates the complex plane of the Mellin moment variable 
$N$ in two half-planes. The contour should be closed to the left complex infinity because the Bjorken variable $x$ is less than 1. It intersects the real axis a bit to the right from the critical exponent 
\cite{Alvarez:2016juq}. Then, in Ref. \cite{Alvarez:2019eaa} we apply a complex diffeomorphism to the integrand of this contour integral in order to transform the latter to another contour integral 
that represents the inverse Laplace transformation of the Jacobian of this complex diffeomorphism. The contour is curved with this diffeomorphism in general but we may always rectify it back to the vertical line. 
This vertical line is the main part of the contour in the the inverse Laplace transformation \cite{Alvarez:2016juq}. The Jacobian of this complex map may be represented in terms of the Mellin-Barnes transformation 
with another vertical line again as the main part of the corresponding integration contour. Then, the vertical line from the contour of the inverse Laplace transformation may be curved and deformed to the shape of Hankel contour.
With such a deformation we take this Hankel contour integral off and gain the Euler gamma function in the denominator of the contour integral corresponding to the Mellin-Barnes transformation 
of the Jacobian which becomes a Barnes integral after this. The Barnes integrals
are contour integrals containing ratios of products of Euler gamma functions in numerators and denominators. With such a procedure we transform the original contour integral to a Barnes integral which is
the standard representation for generalized hypergeometric functions.

\section{The running coupling}

In case when the coupling  runs, complex mappings still may be used to relate the BFKL and the DGLAP  equations. The differential DGLAP equation in the small $x$ limit for the unintegrated 
gluon distribution is  
\begin{eqnarray*} 
u\frac{d}{du} \phi(N, u) = \frac{\alpha(u)}{2\pi}\gamma(N,\alpha(u))\phi(N,u) \Rightarrow  \frac{d}{d\alpha} \phi(N, \alpha) = \frac{\alpha}{2\pi}\frac{\gamma(N,\alpha)}{\beta(\alpha)}\phi(N,\alpha),
\end{eqnarray*} 
where the coupling  $\alpha$ has been chosen as a variable in a usual way instead of the scale $u$ for this equation, $\beta(\alpha) = u\frac{d}{du}\alpha(u).$ Then,    
\begin{eqnarray*} 
\phi(N, \alpha) =  \phi_1(N)\exp{F(\alpha,N)} \Rightarrow  \phi(x,u) = \oint_C~dN \phi_1(N)\exp{ \left[F(\alpha,N) -N\ln{x}\right],} 
\end{eqnarray*} 
where $\dis{\frac{\partial}{\partial\alpha} F(\alpha,N) = \frac{\alpha}{2\pi}\frac{\gamma(N,\alpha)}{\beta(\alpha)}}.$ Now, in order to obtain the BFKL equation via mapping, 
we choose a new variable $M$ so that $M\alpha  = F(\alpha,N).$

In order to transform the Mellin moment of the PDFs back to the Bjorken $x$ space, we try to choose a new integration variable $M$ in  such a way that the integrand is 
as simple as possible. The integrand would be a product of Jacobian of the mapping and of the shape function $\phi_1(N)$ rewritten in terms of the new variable $M.$

\section{Conclusion}

As we have proposed in this talk,  diffeomorphisms in the complex plane of the Mellin moment are an efficient way to relate the DGLAP and the BFKL equations, in both the cases of the fixed or of the running 
coupling. We try to extract more benefits from this observation and adjust the mapping in each case so that  the inverse 
Mellin transformation simplifies by reducing itself to the inverse Laplace transformation which corresponds to the table integrals in many of the cases. 
The new variable should be chosen to reach the maximal simplicity for integrands. However, in any case we should choose these new variables in such a way that at the intermediate step the inverse Laplace transformation of the Jacobian 
of the complex map appears because we need such a structure of integrands of the contour integrals in order to deform their shapes by rectifying them 
or by curving them to the Hankel contours and to represent then these integrals in terms of the Barnes integrals.

\ack{This work was supported by Fondecyt (Chile) grant 1050512 and by DIUBB (Chile) Grant Nos. 102609 and GI 153209/C. 
This paper is based on the talk at ACAT 2019, Saas-Fee, Switzerland, March 10-15, 2019, and the author is grateful
to Andrei Kataev for inviting him. The financial support of author's participation in ACAT 2019 has been provided by DIUBB via Fapei funding.
The author thanks the referee of \cite{Alvarez:2016juq} who has updated the QCD bibliography there.}

\end{document}